\documentclass[electronic, preprint]{vgtc}        

\ifpdf
  \pdfoutput=1\relax                   
  \usepackage{graphicx}                  \DeclareGraphicsExtensions{.pdf,.png,.jpg,.jpeg}
\else
  \usepackage{graphicx}                
  \DeclareGraphicsExtensions{.eps}     
\fi

\graphicspath{{figures/}{pictures/}{images/}{./}} 

\usepackage{microtype}                 
\PassOptionsToPackage{warn}{textcomp}  
\usepackage{textcomp}                  
\usepackage{mathptmx}                  
\usepackage{times}                     
         
\usepackage{cite}                      
\usepackage{tabu}                      
\usepackage{booktabs}                 

\usepackage[font={small,sf},tableposition=top]{caption}
\usepackage{colortbl}
\usepackage{paralist}
\usepackage[online,flushleft]{threeparttable}
\usepackage{makecell}
\usepackage{multirow}
\usepackage{multicol}
\usepackage{color}
\usepackage{gensymb}

\def\techname{DualStream}
\def\usera{Elijah}
\def\userb{Daneel}

\vgtcinsertpkg

\title{ \techname{}: Spatially Sharing Selves and Surroundings \\ using Mobile Devices and Augmented Reality}

\author{Rishi Vanukuru\textsuperscript{1}\thanks{Email: \href{mailto:rishi.vanukuru@colorado.edu}{rishi.vanukuru@colorado.edu}} \and Suibi Che-Chuan Weng\textsuperscript{1} \and Krithik Ranjan\textsuperscript{1} \and Torin Hopkins\textsuperscript{1} \and Amy Banic\textsuperscript{2} \and Mark D. Gross\textsuperscript{1} \and Ellen Yi-Luen Do\textsuperscript{1}\thanks{Email: \href{mailto:ellen.do@colorado.edu}{ellen.do@colorado.edu}}}

\affiliation{\scriptsize \textsuperscript{1} ATLAS Institute, University of Colorado Boulder, USA \\ \textsuperscript{2} Interactive Realities Lab, University of Wyoming, USA}

\shortauthortitle{Vanukuru \MakeLowercase{\textit{et al.}}: \techname{}: Spatially Sharing Selves and Surroundings using Mobile Devices and Augmented Reality}

\teaser{
  \centering
  \includegraphics[width=\textwidth]{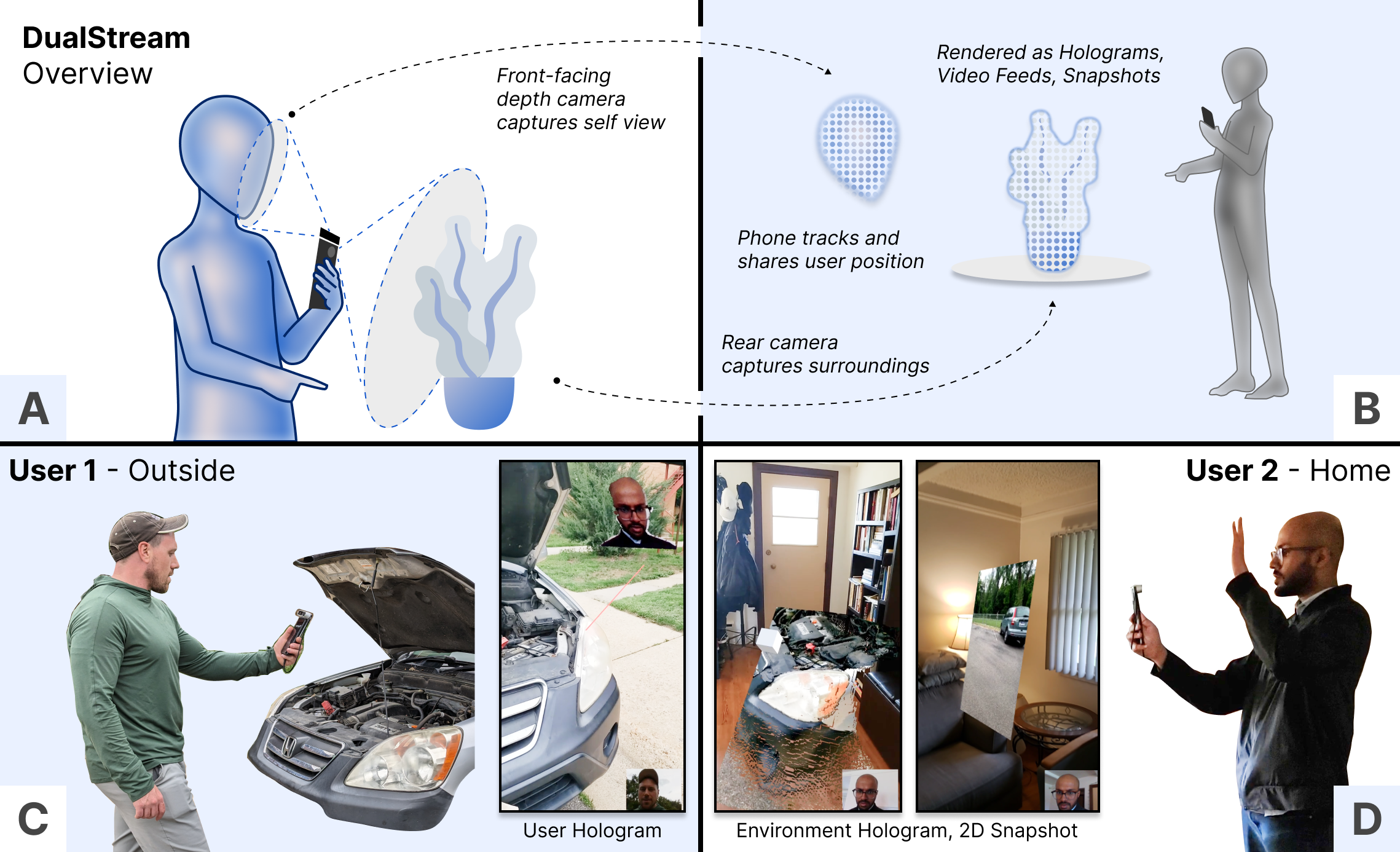}
  \caption{An overview of the \techname{} system. (A \& B) An illustration of how the front and rear cameras of a mobile device can be used to share information about self and surroundings with remote collaborators. The captured information is used to create 3D Holograms, Spatial Video Feeds, and Environment Snapshots. (C) \techname{} being used to share information about a car's engine, while viewing a remote expert as if they were in the same location. (D) The expert viewing the shared 3D hologram of the car, and a 2D snapshot of the surrounding remote environment.}
  \label{fig:teaser}
}

\abstract{
In-person human interaction relies on our spatial perception of each other and our surroundings. Current remote communication tools partially address each of these aspects. Video calls convey real user representations but without spatial interactions. Augmented and Virtual Reality (AR/VR) experiences are immersive and spatial but often use virtual environments and characters instead of real-life representations. Bridging these gaps, we introduce \textit{\techname{}}, a system for synchronous mobile AR remote communication that captures, streams, and displays spatial representations of users and their surroundings. \techname{} supports transitions between user and environment representations with different levels of visuospatial fidelity, as well as the creation of persistent shared spaces using environment snapshots. We demonstrate how \techname{} can enable spatial communication in real-world contexts, and support the creation of blended spaces for collaboration. A formative evaluation of \techname{} revealed that users valued the ability to interact spatially and move between representations, and could see \techname{} fitting into their own remote communication practices in the near future. Drawing from these findings, we discuss new opportunities for designing more widely accessible spatial communication tools, centered around the mobile phone.
}

\CCScatlist{
  \CCScatTwelve{Human-centered computing}{Mixed/augmented reality}{}{};
  \CCScatTwelve{Human-centered computing}{Collaborative Interaction}{}{};
    \CCScatTwelve{Human-centered computing}{Mobile computing}{}{}
}

\begin{document}
\maketitle
\section{Introduction}
 
Human communication is fundamentally tied to our perception of one another in a shared physical environment. We constantly process and add to a three-dimensional (3D) audio-visual canvas of gestures and conversations. Movement is also a key aspect, as embodied cognition \cite{foglia2013embodied} and proxemics shape the way we navigate space in social contexts \cite{kendon1990conducting}. With remote communication becoming increasingly integral to work and personal life, a key challenge is to develop tools that can match the experience of in-person interactions. Video-conferencing
is one of the most common forms of audio-visual remote communication today, and enables synchronous conversations using real-world representations of ourselves and the environments we inhabit. Presenting this information on two-dimensional (2D) screens, however, makes it difficult to establish a shared frame of reference and interact in a more spatial manner \cite{ohara_blended_2011}. Recent advances in Augmented and Virtual Reality (AR/VR) devices, as well as camera technology that can simultaneously capture color and depth information, have resulted in systems where complete human representations and entire physical environments can be streamed and interacted with remotely in 3D \cite{orts-escolano_holoportation_2016, thoravi_kumaravel_loki_2019}. While such experiences may come close to replicating in-person communication, the systems involved are expensive, confined to specific locations, and rely on devices such as head-worn displays that are still relatively unfamiliar to wider audiences. In contrast, mobile AR applications have given millions of people around the world a glimpse of the power of spatially interacting with information \cite{chatzopoulos2017mobile}. 
By integrating the advanced video and depth capture technology present in modern mobile devices, and the spatial interaction capabilities of mobile AR applications, mobile devices are well-positioned to bridge the gap between spatial remote communication and wider availability.

To demonstrate this potential, we develop \textit{\techname{}}—a mobile remote communication platform where users share 3D views of each other and their surroundings, and can interact with these representations spatially in AR (\autoref{fig:teaser}). \techname{}’s hardware comprises a mobile device with an externally-mounted front-facing depth camera. \techname{} captures two ``streams'' of 3D information from the front and rear cameras, corresponding to the user and their immediate surroundings.
Using these streams, we create remote 3D representations of (1) ourselves, which look and move as we do in real life, and (2) our surroundings, that are spatially consistent with their real-world locations. We develop a series of features such as seamlessly moving between fixed and spatial views, and using environment snapshots to create a persistent shared space for remote communication. Thus, \techname{} enables users to simultaneously feel they are ``being there'' in a remote location (by spatially interacting with shared environment views), and that remote participants are ``being here'' in their local environment (as independently moving user representations in space). We demonstrate that unlike fixed volumetric capture setups tethered to single rooms and reliant on expensive hardware, \techname{} can enable the sharing of selves and spaces, anywhere and anytime. \techname{} leverages the familiarity of personal mobile computing, and provides a more spatial and immersive experience than the status quo of video conferencing.

We conducted a formative evaluation of \techname{} with users in the lab and outside, to understand how people compare mobile spatial communication with their current practices of remote collaboration. Findings from this study showcase the potential for \techname{} to enhance everyday spatial communication, and provide insight into areas which require improvement before such experiences become widely adopted. Our key contributions are:
\begin{enumerate}
    \item The development of a mobile-based remote communication platform—\techname{}—that can simultaneously share spatial representations of users and their environments, with features that enable users to combine 2D and 3D representations of self and surroundings,
    \item Real-world scenarios which showcase the applicability of mobile spatial communication to a range of contexts, and
    \item Findings from a formative evaluation of \techname{} in the lab and in the real world, showcasing its potential and highlighting challenges and opportunities for future work.
\end{enumerate}

\section{Related Work}
\techname{} builds upon prior work on AR/VR collaboration, video conferencing, as well as the specific use of mobile devices to support both these forms of communication. We focus on systems that consider more realistic, real-time representations of users and their environments. We also draw on Harrison and Dourish’s distinction between interaction in ``space'' — using direct geometric arrangements or metaphors to make sense of interpersonal communication — and in a specific ``place'' — locations imbued with contextual and social meaning \cite{harrison_re-place-ing_1996, dourish_re-space-ing_2006}.

\subsection{Bringing Real-World Information into Immersive Collaboration}
Immersive collaboration takes place in shared digital worlds via VR or by making use of each user’s local environment via AR. Interaction in space is a necessary characteristic of such immersive systems. However, questions of how best to represent one’s real self, and incorporate aspects of real-world places, still remain.

\subsubsection{Real-World Selves}
Early work by Bailenson et al. \cite{bailenson_effect_2006} demonstrated the value of using realistic representations of collaborators in virtual environments. Many studies have since further highlighted the beneficial impact of realism and fidelity \cite{yoon_effect_2019,latoschik_effect_2017, waltemate_impact_2018} on presence and collaboration. Commercial immersive communication tools such as Spatial\footnote{Spatial: \href{https://www.spatial.io/}{spatial.io}} also offer users the option to create realistic avatars. Another approach is to capture and stream real-time visual information of users. An early system by Billinghurst and Kato involving collaboration via AR headsets \cite{billinghurst_collaborative_2002} overlaid real video feeds of users over fiducial markers. There have been further efforts to integrate video calls with immersive environments \cite{10.1145/3411763.3451546}. Recent improvements in 3D content capture have enabled systems where full-body information can be streamed and reconstructed remotely in real-time \cite{orts-escolano_holoportation_2016, thoravi_kumaravel_loki_2019, gamelin_point-cloud_2021, yu_avatars_2021, langa_multiparty_2022}.

\subsubsection{Real-World Places}
Using 360-degree video capture, VR systems have been able to provide more complete pictures of local environments to remote users inhabiting a fixed perspective \cite{zhang_xrmas_2021}, moving based on local user control \cite{piumsomboon_shoulder_2019}, or independently using telepresence robots \cite{jones_belonging_2021}. With room-scale depth capture, projects such as XSpace \cite{herskovitz_xspace_2022} and Re-locations \cite{fink_re-locations_2022} enable the creation of blended locations for remote, multi-view collaboration. Systems such as Loki \cite{thoravi_kumaravel_loki_2019} and Holoportation \cite{orts-escolano_holoportation_2016} are also capable of streaming environmental depth information. Recent surveys have highlighted the importance of situated spatial collaboration in the context of remote assistance and collaboration on physical tasks \cite{fidalgo2023survey, wang2021ar}.

Taken together, these immersive systems have brought us close to replicating real-world environments and interactions in real-time. However, challenges to depth-based spatial communication such as scalability and transmission reliability are yet to be solved \cite{petkova_challenges_2022}, and the devices required for capture and display are expensive, relatively unfamiliar, and far from widespread use.  

\subsubsection{Mobile Spatial Collaboration}
Mobile devices have proven to be ideal platforms for making immersive experiences available to wider audiences today \cite{chatzopoulos2017mobile, dagan_project_2022}. Mobile AR applications are now commonplace, and have provided millions of people a glimpse of spatial computing via games and social apps like Snapchat\footnote{Snapchat: \href{https://www.snapchat.com/}{snapchat.com}} and Pokemon Go\footnote{Pokemon Go: \href{https://pokemongolive.com/}{pokemongolive.com}}. When considering collaborative mobile AR experiences, research has explored their use in co-located contexts, for games \cite{bhattacharyya_brick_2019}, social activities \cite{dagan_project_2022}, and spatial problem solving \cite{wells_collabar_2020, grandi_design_2018}. Mobile AR has also been used to support remote collaboration \cite{muller_remote_2017, datcu_handheld_2016} in the contexts of lab-based activities \cite{villanueva_colabar_2022} and design critiques \cite{li_arcritique_2022}. Work by Young et al. on mobile telepresence has considered the ways in which mobile devices can share user and environment representations in real-time, first via interaction in three degrees of freedom \cite{young2019mobiletelepresence}, and subsequently using 360 degree cameras mounted on mobile devices in the Mobileportation project \cite{young2020mobileportation}.

\subsection{Making Video Conferencing more Spatial}

Similar to how immersive collaboration incorporates aspects of space by default, video conferencing tools rely on capturing and streaming real-world audio-visual information of remote participants in real-time. Prior research has studied the impact of screen-based spatial metaphors on the sense of space and co-presence \cite{hauber_spatiality_2006}. These metaphors include viewing remote video feeds around conference tables \cite{vertegaal_gaze_1999}, and enabling video feeds to move in and out of virtual rooms to indicate informal and formal meeting spaces \cite{panda_alltogether_2022}. While these systems offer a strong virtual frame of reference, the visual information is still limited to being viewed on fixed 2D screens, and the resulting sense of spatial interaction is inferred rather than direct. One way in which video conferencing has been made more spatial is by having displays and cameras that literally move in space. Through cameras controlled by remote users \cite{ranjan_dynamic_2007}, robotic-arm-mounted displays that mirror head gestures \cite{sakashita_remotecode_2022}, and telepresence robots \cite{li_asteroids_2022, neustaedter_being_2018}, studies have investigated how spatial cameras and displays can impart a greater sense of co-presence and a common shared reference frame. However, like the more advanced AR/VR collaboration systems discussed earlier, these spatial video conferencing setups are also cumbersome, expensive, and difficult to scale to more general communication. In contrast, mobile devices have brought video calling to the masses. Mobile video calls support informal ``small talk'' interactions, task-oriented functional conversations, as well as ``show and talk'' interactions where real-world objects are the focus \cite{ohara_everyday_2006}. The presentation of spatial information on fixed 2D screens, however, presents challenges around the asymmetries of control, participation, and awareness \cite{jones_mechanics_2015}.

\subsection{What \techname{} does Differently}
 Through \techname{}, we enable users to share representations of their selves and surroundings, and interact with remote collaborators in a spatial manner. Thus, \techname{} replicates features provided by more elaborate volumetric setups \cite{orts-escolano_holoportation_2016, thoravi_kumaravel_loki_2019} and goes beyond existing mobile AR systems by focusing on the spatial sharing of both self and place-based information. A key difference between \techname{} and prior systems such as Mobileportation and ARCritique \cite{young2020mobileportation, young2019mobiletelepresence, li_arcritique_2022} is the ability to share 3D information about oneself. \techname{} offers more control over what is being shared, and enables users to freely combine 2D and 3D information depending on the context and level of detail required. \techname{} also employs a fundamentally different scheme of capturing, compressing, streaming, and rendering depth information. We colorize and compress two streams of depth simultaneously (self and surroundings) and stream them using conventional video networks. The freedom to use \techname{} anywhere and anytime presents possibilities not available in immersive AR/VR systems. Further, we lean into the familiarity of mobile devices and videoconferencing, by enabling users to seamlessly switch between spatial AR calls and screen-based video calls. By presenting information about self and shared places in AR, \techname{} helps provide a strong frame of reference for inherently spatial media. The ability to visualize video feeds in 3D enables \techname{} to simulate the notion of displays moving in space, as employed by related work on spatial video calls \cite{sakashita_remotecode_2022}. When taken together, \techname{} enables spatial remote communication in personal spaces, truly remote assistance, and supports the creation of blended spaces for interaction.

\section{The \techname{} System}
\label{section:system}

\techname{} supports \textbf{spatial interaction} via mobile AR, and enables users to share various representations of \textbf{self} and \textbf{place} (their local surroundings) with remote collaborators. The implementation of this prototype consists of a mobile device 
with an externally mounted, front-facing depth camera (Intel RealSense D435) directly connected to the device via a USB-C cable. Although some modern mobile devices are already equipped with front and rear-facing depth cameras, it is not yet possible to access depth data while also capturing color video and running AR experiences. Therefore, we use the externally mounted camera to explore capabilities that will likely be achievable by on-device solutions in the near future. Even in this form, the phone and depth camera setup is not tethered to a specific location, and is more readily available than complex volumetric capture systems involving AR/VR headsets and multiple cameras.

\subsection{Establishing a Shared Spatial Frame of Reference}
\label{system:referenceframe}
When starting a call using \techname{}, users must first scan horizontal surfaces in their local environment to help the phone track its own position and orientation. Users then touch an appropriate location on the screen to place the ``anchor'' object (represented as a small blue cube) for the shared room. The position and orientation of the user's phone relative to the anchor is then streamed to any remote participants. This data helps to consistently position the shared user and environment representations.

\subsection{Sharing Realistic Representations of Self}
Once users have set up the shared reference frame, they can use the buttons on the interface to begin sharing representations of themselves. These representations are placed and moved based on user movement, thereby creating remote representations that look and move the way we do. The two main user representations in \techname{} are (1) \textbf{3D Hologram}, where the local user sees a point-cloud representation of the remote user's face, and (2) \textbf{Spatial Video}, where the front-facing camera feed of the remote user's head and shoulders is displayed in real-time, with or without the remote background (\autoref{fig:teaser}C). As with other video-conferencing applications, users can see their self-view in the form of a small video in the bottom-right corner of the screen. Users have complete control over which representation of themselves is shared with remote participants, and can also turn off their representation, displaying a small white cube instead. These representations enable users to feel like the person they are talking with is in the same space as they are.

\subsection{Sharing Real-world Environments}
Users can also share their environments in real time. Similar to the self representations, users have full control over what environment representation they share remotely. \techname{} offers two options for sharing environment information (\autoref{fig:teaser}D):

\begin{enumerate}
    \item \textbf{Environment Hologram}: A 3D representation of the environment is projected as a point cloud from the position of the remote user, to appear to be situated in the local space. As the remote user moves around, this point cloud also moves to represent the changing surroundings. 
     \item \textbf{Environment Video Feed}: A user shares the view of their immediate surroundings as seen on the phone screen. The local user sees a rectangular frame with this video feed attached to the remote user at a fixed distance from their self-representation, mapped to the perspective and field of view of the remote user. As the remote user moves, this rectangular frame moves around with them, functioning as a portal into the remote space.
     This representation is useful for sharing parts of the environment that are primarily 2D in nature or distant objects whose depth cannot be accurately captured.
    \end{enumerate}

\subsection{Key Features}
\techname{} consists of a basic user interface with a row of buttons in the bottom-left of the screen corresponding to the user and environment representations. By pressing these buttons, a secondary set of options appears, allowing users to switch between representations, take snapshots, activate the front-facing depth camera, and re-position the anchor object (\autoref{fig:scenarios}L).

\subsubsection{Environment Snapshots}
Users can only share real-time information about the environment directly in front of themselves, as viewed by the phone's rear camera. \techname{} enables users to take \textbf{snapshots} of the environment (as a video frame or hologram) in its current state. These snapshots then persist in the location where they were taken in the shared remote space. This can be used to freeze video frames or holograms in place for later discussion without the live feed (\autoref{fig:teaser}D). By taking multiple snapshots, users can share a greater amount of their local environment in a spatially persistent manner. Once a user places a snapshot of their environment in their collaborator's view, they see a spatially anchored annotation indicating the location from where the snapshot was taken (as a small cube), and the visual area that was covered by the snapshot (as a semi-transparent plane). This helps provide feedback about previously shared information and aids in the creation of multiple snapshots of continuous scenes. These annotations and snapshots can be selectively displayed or hidden, based on the context of the conversation.

\subsubsection{Pointing}
Users can point at remotely shared environments and snapshots by touching the phone display. This triggers a laser pointer that the remote user can view in their own local space. The pointer helps approximate a basic deictic gesture and aids with spatial referencing in the shared environment (\autoref{fig:teaser}C).

\subsubsection{Switching from AR to Screen call}
\techname{} allows users to quickly and seamlessly switch from a spatial AR call to a conventional screen-based video call (\autoref{fig:scenarios}B) using buttons located below their self-view in the bottom-right of the screen (\autoref{fig:scenarios}L).

\section{Implementing \techname{}}

\techname{} is implemented as an Android application, and can function on any Android device capable of supporting AR experiences via Google's ARCore API\footnote{Google AR Core \href{https://developers.google.com/ar}{developers.google.com/ar}}. Across the development and formative evaluation, we used Samsung Galaxy S9 devices. We used Unity 2022.2\footnote{Unity: \href{https://unity.com}{unity.com}} to develop and build this application. In order to fully capture and stream both the ``dual'' streams, a front-facing depth camera (Intel RealSense D435\footnote{Intel RealSense D435: \href{https://www.intelrealsense.com/depth-camera-d435/}{intelrealsense.com/depth-camera-d435/}}) is mounted on top of the Android device. \techname{} does not need this external camera to function, and can be used to view representations of remote users and their environments, as well as share local environments remotely, even without the depth camera.

\subsection{Local Data Capture}
\subsubsection{Self}
We use the RealSense camera to simultaneously capture frames of Color (RGB24 format) and Depth (Z16 format). When the phone is held at a comfortable viewing distance (approximately 0.4 meters away from the face), this captures the user's head and upper torso. We use the RealSense SDK for Unity\footnote{RealSense SDK: \href{https://www.intelrealsense.com/sdk-2/}{intelrealsense.com/sdk-2}} to access this data within the \techname{} application. The large size of depth buffers makes it bandwidth-intensive to transmit in real-time over conventional networks. Therefore, we convert the depth buffer into a color image, where the color value of each pixel corresponds to the depth at that point. We use the colorization process within the RealSense SDK, and tune it such that depth information of up to 0.8 meters away from the camera is encoded in the entire color range to ensure maximum depth resolution retention. We also align the resultant color and colorized depth images to represent the same visual area (accounting for differences in field of view).

\subsubsection{Surroundings}
We access the rear camera feed of the mobile device using the ARCore SDK for Unity. The ARCore Depth API can provide estimates of environment depth, which is inferred using image processing and sensor fusion \cite{du_depthlab_2020}. Inspired by the colorization approach in the Intel RealSense SDK, we convert each depth frame obtained from ARCore into a single colorized RGB frame, where depth is encoded in the pixel hue using the Turbo color mapping scheme \cite{mikhailov2019turbo}. We process the colorized depth frames to encode distances of up to 2 meters away from the device.

\begin{figure}[h]
  \centering
  \includegraphics[width=\linewidth]{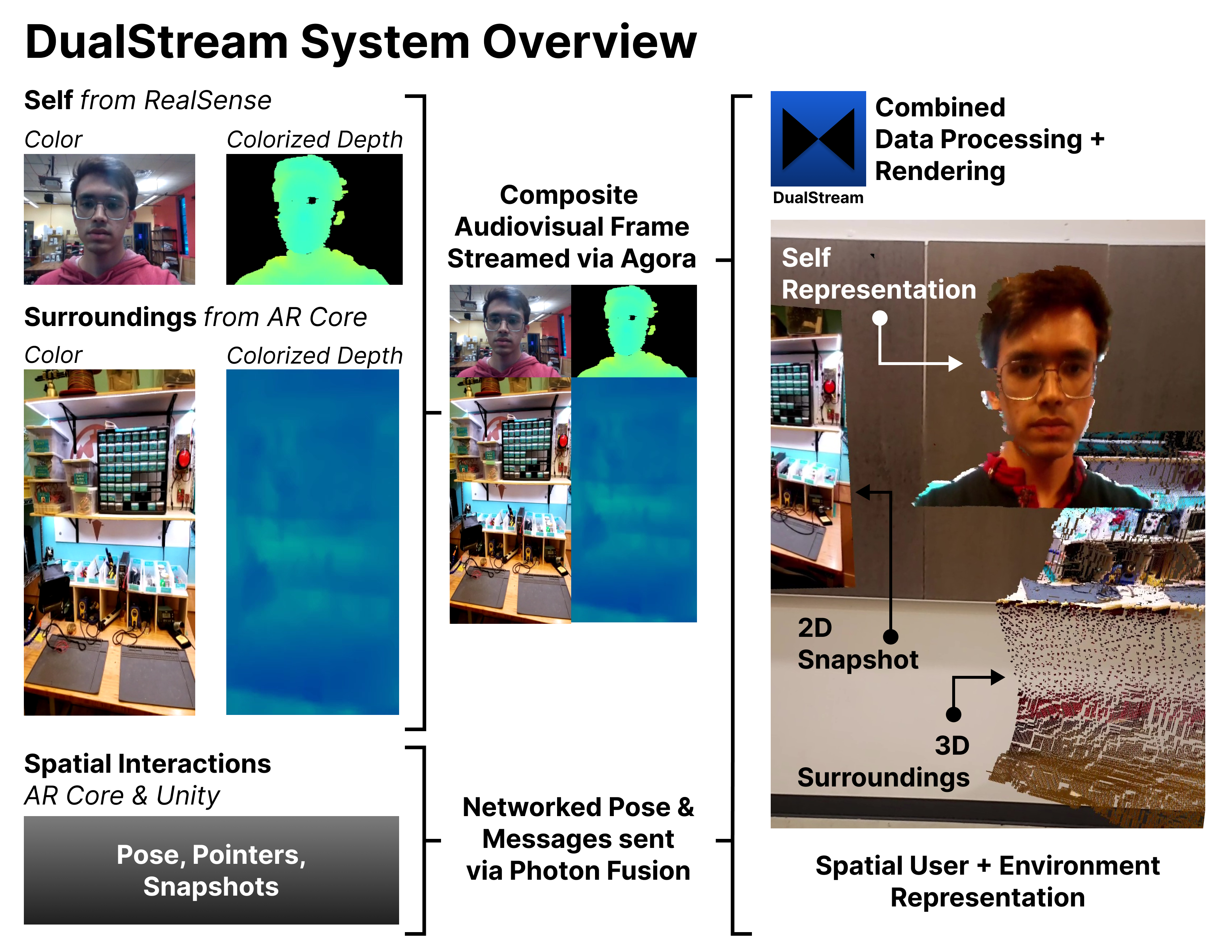}
  \caption{Implementation overview of \techname{}: System diagram of data capture, processing, streaming, and rendering. Audio-visual information is streamed via Agora, while other messages relating to movement and snapshots are streamed via Photon Fusion.}
  \label{fig:implementation}
\end{figure}

\subsection{Networking \& Rendering}
When users launch the \techname{} application, they automatically join a shared room and can begin streaming information to other remote participants. During testing, we used \techname{} to support communication between up to four simultaneous users.

\subsubsection{Position and Interaction Data}
To stream the mobile device pose and other messages that handle user interactions, we use the Photon Fusion SDK for Unity\footnote{Photon Fusion: \href{https://www.photonengine.com/}{photonengine.com}}. A custom script synchronizes position and orientation relative to the anchor object in each user's local frame of reference. Interaction messages (such as changing the type of user and environment representations, pointing, taking snapshots) are also handled via Fusion. Photon is optimized for sharing position and interactions with low latency. However, it is not suited for real-time video, and so we use a different mechanism for audio-visual streaming.

\subsubsection{Audio-Visual Data}
We use the Agora Real-time communication SDK for Unity\footnote{Agora Real-time Communication: \href{https://www.agora.io/en/products/video-call/}{agora.io}}, which provides the ability to stream audio and video information over managed cloud servers. Audio is captured from the device microphone and streamed in a single channel. To efficiently share color and (colorized) depth information from both front and rear cameras, we combine the four individual color frames into a single composite frame, and stream this composite frame via a custom camera capture software device (see \autoref{fig:implementation}). This also ensures that all streams of individual color and depth information are synchronized across the network, with an average latency of under one second.

\subsubsection{Remote Depth Rendering}
Once a remote composite frame is received, we extract the individual color and colorized depth frames, and pass them through custom shaders that generate various representations of the user and their environment. The shaders take into account the camera field of view and maximum depth captured in order to render accurate representations of real-world information. They are also optimized to render representations that appear like point clouds but are less computationally intensive, allowing for a smooth experience on mobile devices. The quality of the 3D environment is lower than that of the 3D user hologram. This is because although the RealSense captures real depth information, the ARCore Depth API only provides a best estimate at the time. However, we assessed that the quality of both 3D renderings was sufficient for a proof-of-concept, in order to get initial feedback from users. The user representations could convey essential facial contours, and the environment representations provided enough depth to distinguish between horizontal, vertical, and curved surfaces. Despite the many parallel computational processes involved, \techname{} runs consistently at 30 frames per second on the Samsung Galaxy S9.

\section{\techname{} in Action}
\label{section:scenarios}

\begin{figure*}[h!]
  \centering
  \includegraphics[width=\textwidth]{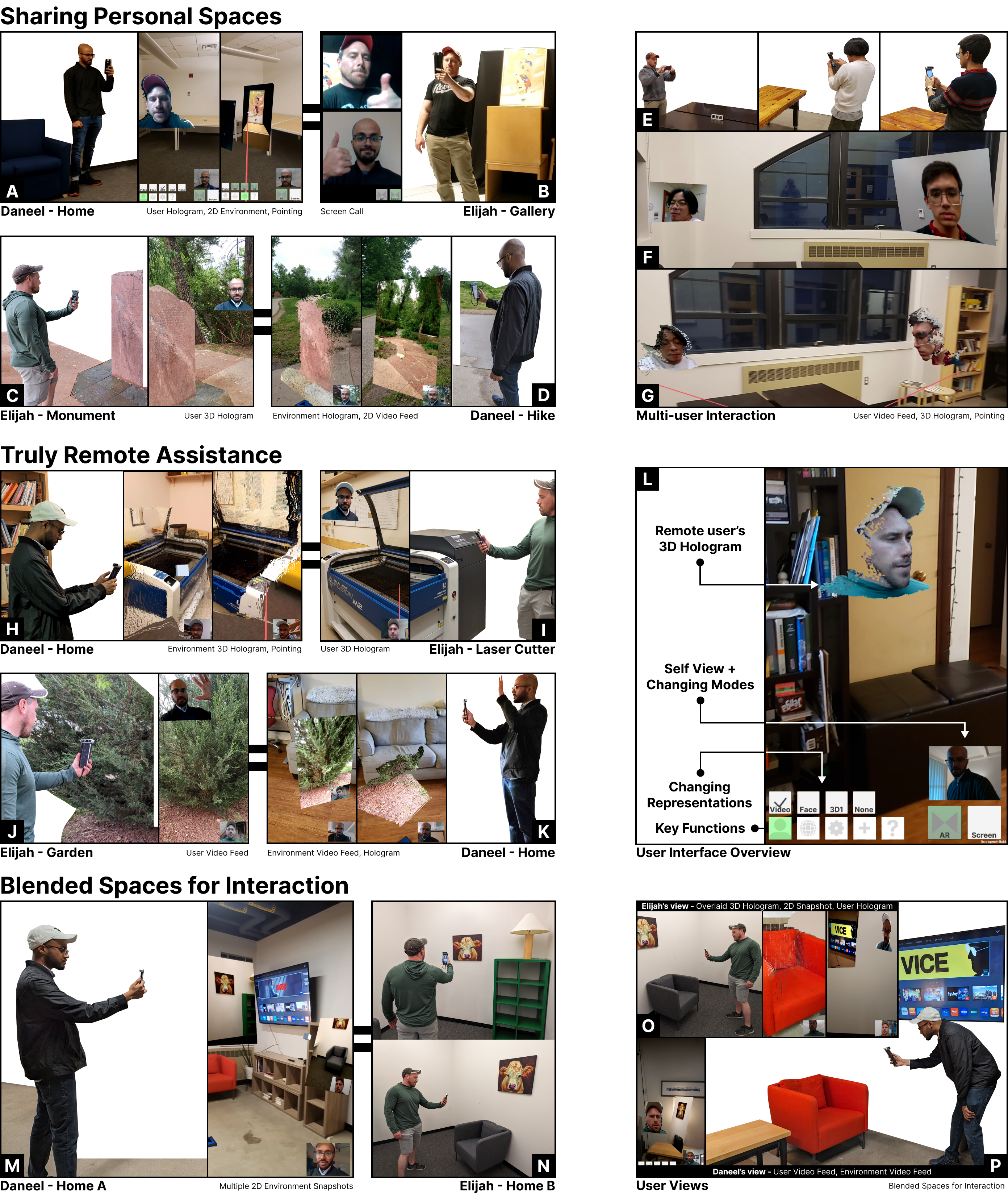}
  \caption{\techname{} System Scenarios: (Top) Examples of \usera{} and \userb{} using \techname{} to share personal spaces, while one user is in an art gallery (A, B), on a hike (C, D), or discussing with multiple members in a team (E, F, G). (Middle) \techname{} being used for remote assistance in a range of contexts - to learn how to work with a laser cutter (H, I), or seeking advice on gardening (J, K). Sub-figure L consists of an overview of the user interface. (Bottom) Demonstrations of how \techname{} creates blended spaces for interaction by capturing moments of space in time (M, N): working with different configurations of the same physical painting, and enabling both users to dynamically share and blend information about screens and furniture from local and remote spaces (O, P).}
  \label{fig:scenarios}
\end{figure*}
 
The core value of \techname{} comes not from replicating cutting-edge volumetric collaboration systems (e.g., \cite{thoravi_kumaravel_loki_2019, orts-escolano_holoportation_2016}), but rather from its ability to enable interaction scenarios that are difficult to achieve via immersive AR/VR systems today. The freedom of using a system based on widely available devices and not tethered to a specific physical location provides unique opportunities. We now demonstrate some examples of the various remote communication contexts that \techname{} can support, using scenarios where two friends — \usera{} and \userb{} — interact with each other. Each example in this section is a result of the research team brainstorming, implementing, and walking through scenarios that highlight the breadth of possible configurations of use. The figures representing each scenario are all taken from screen captures of \techname{}, with further footage available in the supplementary video figure.

\subsection{Sharing Personal Spaces}
While mobile video conferencing helps show places to remote viewers, \techname{} enables users to bring friends and family into their homes from afar, and take them along during travels near and far. For example, \usera{} can call \userb{} via \techname{} from an art galley, and share views of a painting with \userb{} at home. They can also revert to a screen-based video call at anytime, whenever the spatial component of the conversation is completed (\autoref{fig:scenarios} A, B). Both \usera{} and \userb{} could be hiking outdoors, and can use \techname{} to have a quick chat to share views of monuments, trees, and the creek (\autoref{fig:scenarios} C, D). \techname{} can also support larger spatial gatherings. For example, instead of conducting a daily check-in meeting over desktop video-conferencing, \usera{} and their teammates could join a quick spatial call via \techname{}, and appear to all be present around each other's workspaces (\autoref{fig:scenarios} E, F, G). Once everyone has shared their updates, the members can return to mobile or desktop-based video calls for longer discussions.

\subsection{Truly Remote Assistance}
\techname{} can potentially enable people to seek spatial remote assistance from anywhere given a mobile device and internet connection. For example, \usera{} could ask \userb{} for help in fixing the battery of their broken-down car by sharing a 3D hologram that \userb{} can view and point at from their home (\autoref{fig:teaser} C and D). \usera{} can also see \userb{} spatially move around them. Similarly, \usera{} could share a 3D hologram of a laser cutter that they are setting up, to give a remote walk-through to \userb{} who will work on the machine soon (\autoref{fig:scenarios} H, I). The option to capture and stream the environment video feed enables \techname{} to function outdoors as well, and share video information of distant objects that depth cameras cannot capture. For example, \usera{} can share 3D close-up views of the soil, and 2D views of the tree when seeking gardening advice from \userb{} (\autoref{fig:scenarios} J, K). 

\subsection{Blended Spaces for Interaction}
Unlike existing work in which environment sharing is determined by the system, \techname{} gives users agency to construct their own blended, shared place. This enables users to combine and overlap remote environments. For example, \usera{} and \userb{} can share 2D snapshots of objects in each others' rooms such as paintings and a TV, while also sharing a 3D hologram of a piece of furniture they plan to swap. \usera{} can then see the remote couch overlaid on the real couch in their own space (\autoref{fig:scenarios} O, P). This agency enables users to capture not just space, but moments of time in space. \usera{} can take a snapshot of a painting, move it to another wall, and take a second snapshot. In \userb{}'s space, there now exist two copies of the painting, which they can use to advise \usera{} on which arrangement looks better (\autoref{fig:scenarios} M, N).
\section{Formative Evaluation}
We conducted a formative evaluation of \techname{} to (1) gain feedback on the experience of interacting with remote collaborators via \techname{}, and (2) identify key areas of improvement for the prototype. The approach of this formative evaluation was inspired by the notion of Experience Prototyping \cite{buchenau2000experience}. \techname{} in its current form is therefore designed as a probe to enable \textit{``... others to engage directly in a proposed new experience''}, and \textit{``... provide inspiration, confirmation or rejection of ideas based upon the quality of experience''}. This evaluation took place in the lab and remotely, and each study session lasted between 45 minutes and 1 hour. The study procedures were approved by our university's institutional review board (CU IRB Protocol 23-0035). 

\subsection{Participants} 
To evaluate \techname{} in a controlled environment, we put out an open call for participants in our university community. Of those who responded, we excluded members who were already aware of the prototype through demonstrations of earlier versions, and thus we recruited five in-person participants. We also invited five participants from outside our city to use \techname{} from their locations and engage in a remote call with a researcher. These participants were known to the researchers, but were unaware of this project and had not used the prototype before. We used the remote evaluation as a means of validating \techname{} in real contexts where such a prototype is likely to be used. Details about the participants, their familiarity with AR/VR technology, and usage of mobile remote communication tools can be found in \autoref{tab:participants}.

\subsection{In-lab Evaluation}
\begin{figure}[h]
  \centering
  \includegraphics[width=\linewidth]{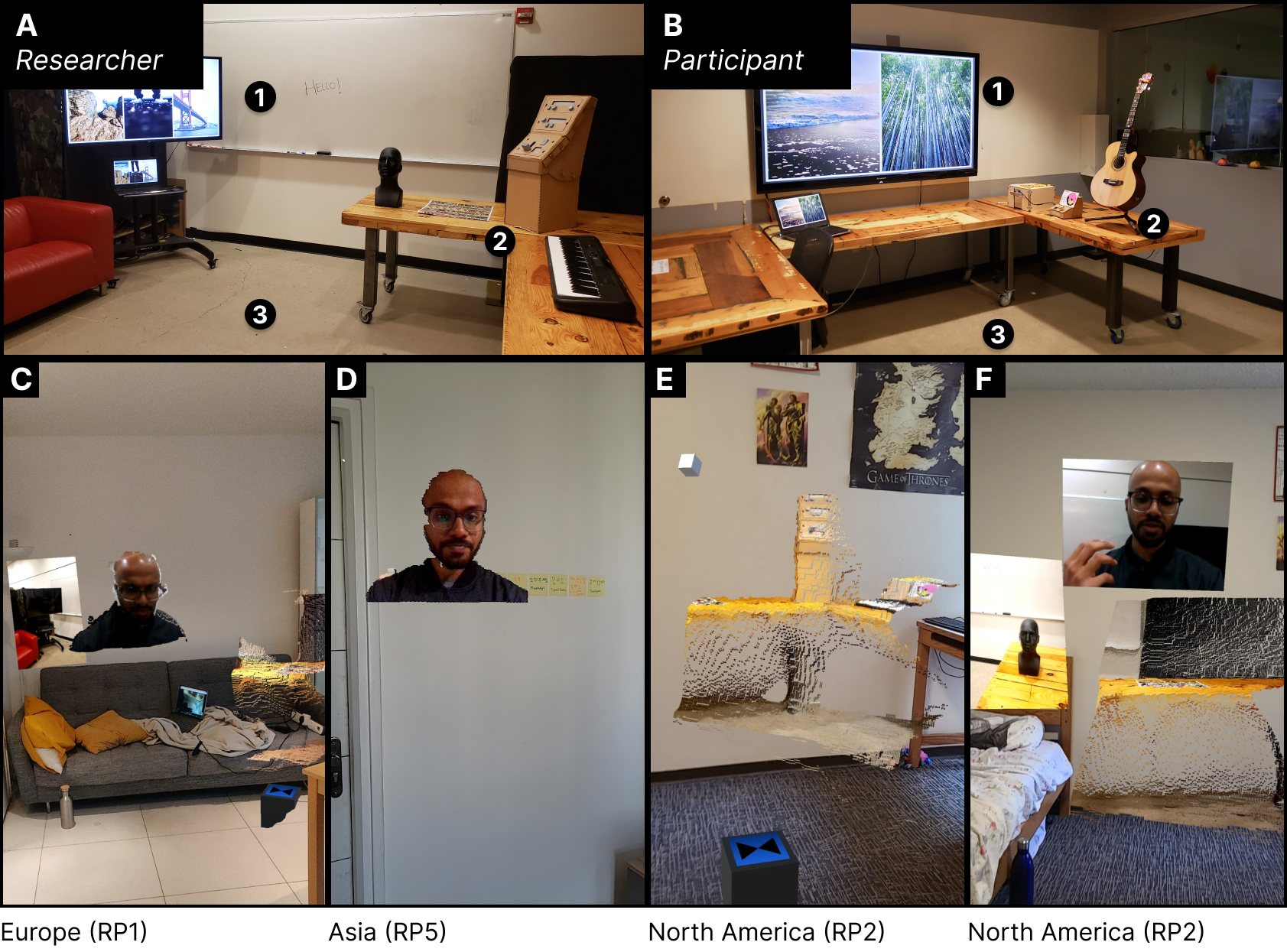}
  \caption{ (Top) An overview of the in-lab study setup. Participants were in location B, while the researcher was in location A. Locations were arranged to provide sufficient room to move and view remote content (3). Both spaces had a range of media such as (1) whiteboards and displays, and (2) 3D objects. (Bottom) A collection of screen views from the remote study (C, D, E, F).}
  \label{fig:studysetup}
\end{figure}

The session began with an introductory discussion where the participant and researcher were in the same physical lab space. The participants answered questions about their usage of remote communication tools and experience with AR/VR technologies. The participants were then directed to a separate lab space, where the researcher helped them start using \techname{} by placing the anchor object and briefing them about the various features of the application. The researcher then returned to the original lab space and joined the call. Both lab spaces were controlled and had a variety of 2D and 3D media to enable a rich range of possible shared content (\autoref{fig:studysetup} Top). Both the participant and the researcher had a mobile phone with an external front-facing depth camera, enabling the bi-directional sharing of self and surroundings. Once in the \techname{} call, the researcher demonstrated the various user representations, environment-sharing capabilities, and interactions. The participants were then asked to try the same features. This demonstration took between 15 and 20 minutes. The participant and researcher returned to the first lab space and engaged in a semi-structured interview regarding the participant's experience with the prototype, seeking feedback for future development. We recorded audio and took notes during this interview, which lasted between 15 and 30 minutes.

\subsection{Remote Evaluation}
The \techname{} application was first sent to each participant, along with instructions on how to install it on their phones. The researcher and participant then joined a common video call where the researcher conducted a similar initial interview and briefing as the in-lab sessions. The researcher walked the remote participant through the process of starting \techname{} and placing the anchor object. After this,
the researcher joined the shared experience using \techname{} running on a mobile phone with an attached depth camera. The participants did not have an external depth camera and were thus unable to share self-representations. The rest of the experience was functionally identical to the in-lab study, with the concluding interview taking place over the video call. \autoref{fig:studysetup} (bottom) shows examples of the screen view of remote participants during the study session (included here with permission). Despite the large distances between study sites, the \techname{} application was able to function smoothly without much latency. The screen views also demonstrate that the quality of rendering at remote sites was comparable to what was achieved in controlled settings.

\begin{table}[tb]
\small
\begin{tabular}{
  >{\raggedright}
  p{2cm}
  >{\centering}
  p{0.75cm}
  p{0.5cm}
  p{1.25cm}
  p{2.25cm}
  }
       \toprule 
 \textbf{Participant} & \textbf{Gender} & \textbf{Age} & \textbf{AR Exp.} & \textbf{Mobile Comm.}\\\midrule

P1  & F & 30  & Limited & Daily\\
P2 & F & 32  & Limited & Daily\\
P3 & M & 28  & Familiar & 2-3x / week\\
P4 & M & 24  & Limited & 4-5x / week\\
P5 & F & 28  & Limited  & 2x / week\\
RP1 (Europe) & M & 26  & Familiar & 1x / week\\
RP2 (N. America) & M & 27  & Familiar & Rarely\\
RP3 (N. America) & F & 22  & None & Daily\\
RP4 (N. America) & M & 22  & None & Rarely\\
RP5 (Asia) & F & 26  & Limited  & 2-3x / week\\   
   
\bottomrule
\end{tabular}
\caption{Participants' (In lab - P and Remote - RP) experiences with Mobile AR (AR Exp.) and mobile video communication applications (Mobile Comm.). Mobile AR experience is categorized as \textit{none} (rarely used), \textit{limited} (used a few times at showcases), \textit{familiar} (used before for personal use).}
\label{tab:participants}
\vspace{-10mm}
\end{table}

\subsection{Results and Findings}
After completing all sessions, two members of the research team qualitatively analyzed the notes and interview transcripts. After independently coding the data into various categories, the researchers discussed their findings and compiled a final set of observations based on overarching themes. These themes relate to the perception of users and environments, interactions in shared spaces, areas for improvement, and potential contexts of future use. 

\subsubsection{The Perception of User Representations}
Across users, the spatial movement of the researcher's self-representation increased engagement, presence, and strengthened the feeling of sharing space. Reactions to the user representations themselves were mixed. On the positive extreme, RP5 mentioned that the moving 3D hologram almost made them believe that the researcher was visiting their home for the first time. 
The negative impact of the ``uncanniness'' of the 3D representation was discussed by several participants (P1, P2, P3, P5). However, P5 mentioned that the point cloud aesthetic helped make the 3D representation feel less uncanny, and suggested that such abstractions could be used to ease this feeling. 
P1, P2, P4, and P5 indicated that a more full-body representation would be desirable.
P1 and RP4 mentioned that the spatial video without background was the most useful for them, as it retained a high enough resolution while also removing overlapping information about the researcher's background. In contrast, RP2 said they would prefer to use the full user video feed, as it was familiar and reminded them of traditional video calls. 

\subsubsection{On the Persistence of Environment Objects}  
Many participants felt the persistence of objects (as captured by snapshots) helped anchor conversations in space. P1, P2, P3, and RP2 mentioned that capturing 3D objects in snapshots could enable users to continue referring to shared objects throughout conversations. RP2 spoke of how \textit{``snapshots enable you to move on with the conversation and still quickly refer back to a point''}, while P2 mentioned that \textit{``you can take a snapshot, continue with conversations and not forget about the object that is still there''}. P1, P4, and RP2 noted how the ability to build scenes using multiple snapshots helped in constructing a mental map of the remote space. 
P1 also mentioned that they could see people using this feature to take snapshots of other members of the family as they moved around the house, bringing in the potential for multi-user interactions around a single device. RP2 stated that the act of following the researcher while they were sharing their environment significantly added to their level of engagement and spatial comprehension.

\subsubsection{Comparing 2D and 3D Surroundings} 
While the utility of 3D environment representations was clear to most users, in its current form, many preferred the spatial environment video feed (P1, RP1, RP2), and mentioned that it was sufficient to gain a sense of the relative positions of objects. P2 mentioned practical reasons for using 2D content, stating \textit{``if I’m on a beach or large landscape, 2D is better than 3D because I can’t share the ocean''}. P5 discussed using 2D video for sharing a large environment if the remote user is in a small space, or perhaps \textit{``scaling down the environment''}. Regarding 3D representations, P3 mentioned that \textit{``though rendering could use improvement, it really felt like the objects were in my space''}, and this sentiment was expressed by many other participants. P2 also frequently navigated ``out'' of the shared space by switching to the screen-based call. They mentioned that viewing 3D content all the time was overwhelming, and they appreciated the option to move to the screen-based call because it was familiar.  

\subsubsection{Areas for Improvement}
Concerning the self representations, users mentioned the need for more context about the body (P1, P2, P5), but also cautioned that a combination of a virtual body and a real face might make the experience feel even more uncanny (P5). Spatial constraints and sharing of vastly different environments posed issues, with P5 suggesting that the application should be able to scale the shared environment down to fit smaller spaces. All participants mentioned that the rendering of 3D information required improvements to be truly usable.  
The user interface was minimal, and participants often required instructions from the researcher to use certain functions. Several participants (P1, P2, P3, P5) suggested that the AR content should be editable. The amount of persistent AR content was overwhelming in some cases, and the need to delete unused snapshots became apparent. 

\subsubsection{Future Applications: Practical and Playful}
Many users mentioned that they could see themselves using \techname{} to share new spaces and view them remotely (P1, P3, RP1, RP2, RP5), in the context of looking for apartments to move into for example. Architectural planning was also indicated as a potential context, with P2 and P5 discussing how sharing scaled snapshots of environments between indoor and on-site locations would be particularly helpful. Some users considered contexts of play, by proposing shared escape rooms (P5), bringing in interactive AR content (P1, P4), and adding playful interactions to snapshots, such as walking to trigger actions or moving between persistent scenes (P1). 
\section{Discussion}

\subsection{The Promise of Mobile Spatial Communication}
In developing \techname{}, we were primarily concerned (like many contemporary projects in the aftermath of the COVID-19 pandemic) with adding to the experience of ``being there'' \cite{bly_media_1993}. This objective naturally led to incorporating more realistic spatial representations of each other. In doing so, we uncovered interactions that empower users to not only share, but blend their surroundings in ways that go ``beyond'' in-person interaction \cite{hollan_beyond_1992}. These interactions have analogues in the space of immersive collaboration. Projects have discussed ways to merge \cite{herskovitz_xspace_2022, wang2023scenefusion}, duplicate \cite{yu2022duplicated}, and remix \cite{lindlbauer_remixed_2018} immersive environments. Some of these ideas also emerged from participant reactions to the \techname{} prototype. Participants wondered if shared environments could be scaled, edited, and invited to merge in ways that would foster personal connections.
Future work can explore how such concepts and ideas from immersive AR/VR can be adapted in mobile contexts. Mobile Spatial Communication is also uniquely positioned to act as a bridge modality for cross-reality systems. Applications such as Mozilla Hubs already allow users to join shared virtual spaces using personal computers and AR/VR headsets. By incorporating features from \techname{} into these systems, mobile devices can help users to better navigate the transition between the desktop computing paradigms of today, and the immersive computing environments of the near-future.

\subsection{The Potential for Machine Learning}
Ongoing work in the areas of machine learning and computer vision can potentially help improve the fidelity and completeness of representations created by \techname{}. Systems such as Pose-on-the-Go \cite{ahuja_pose---go_2021} already leverage mobile sensors to predict human pose and gestures. Combining this full-body information with real-time facial capture would be a step towards more complete user representations. Recent work on creating photo-realistic avatars using smartphone cameras \cite{cao2022authentic} demonstrates how mobile devices might soon be capable of computing and rendering high-quality faces for applications in AR/VR communication. Integrating methods for 3D reconstruction such as Neural Radiance Fields is another promising direction, with steps taken towards mobile device-based capture and processing of high-quality environment renderings \cite{rojas2023re, park_instantxr_2022}. While these approaches generate realistic faces and spaces, there needs to be a balance between information that is streamed as-is (like \techname{}) or computed, based on the context at hand.

\subsection{Limitations}
\techname{} is a proof-of-concept of capabilities that are likely to be available on mobile devices in the near future. Throughout the design and development process, we made decisions that prioritized wider access, such as choosing cloud-based networking services to reach a broader audience outside the lab. There are potentially other combinations of present-day hardware (using iOS devices with depth cameras, or local networking setups) that might yield more optimized systems for in-lab studies. While the quality of 3D representations can be improved, testing \techname{} at this stage revealed how even 2D information, when presented spatially in AR, might suffice for specific tasks. Our formative evaluation provides initial insights into how users interact with \techname{} and the contexts in which they might want to use it. The study helped us understand which of the many features we should prioritize, while also uncovering some unexpected and interesting findings. The open-ended nature of the session was instrumental in achieving this. However, the lack of a concrete task makes it difficult to comment on the more immediate utility of \techname{}.
We intend to conduct follow-up evaluations focused on specific aspects of the prototype (e.g., user representations, pointing and gestures) in applied contexts, in order to gain a deeper understanding of how such a system can be designed to best support remote collaboration.

\section{Conclusion}
\techname{} explores how mobile devices can better support spatial remote collaboration. By leveraging the combination of cameras and sensors on mobile phones, as well as their ubiquity relative to AR/VR headsets, \techname{} enables the sharing of spatial representations of both self and place, anywhere and anytime. We discuss the benefits of \techname{} through scenarios and examples showcasing how it can support the sharing of personal spaces, truly remote assistance, and interaction in blended spaces. Findings from a formative evaluation conducted in the lab and in the real world indicate that users found value in real representations, spatial interactions via mobile AR, and the ability to construct persistent shared spaces. Our work lays out new directions for enabling more widely accessible spatial communication tools with mobile devices as the focus.
\acknowledgments{
The authors would like to thank Greg Phillips, Alvin Jude, Gunilla Berndtsson, Per-Erik Brodin, Amir Gomroki, and Per Karlsson for their support and advice throughout this project. We also wish to thank Sandra Bae and Ruojia Sun for their assistance and many helpful discussions. This project was supported by a grant from Ericsson Research, and by the National Science Foundation under Grant No. IIS-2040489.}

\end{document}